\documentclass[12pt,preprint]{aastex}
\usepackage{graphicx}
\usepackage{apjfonts}

\usepackage{lscape}

\shorttitle{ACCRETION DISK EVOLUTION OF BLACK HOLE AND NEUTRON STAR X-RAY BINARIES}

\shortauthors{Authors}

\begin{document}

\title{COMPARING THE ACCRETION DISK EVOLUTION OF BLACK HOLE AND
NEUTRON STAR X-RAY BINARIES FROM LOW TO SUPER-EDDINGTON LUMINOSITY}

\author{Shan-Shan Weng,\altaffilmark{1}  Shuang-Nan Zhang,\altaffilmark{1}}
\altaffiltext{1}{Key Laboratory of Particle Astrophysics,
Institute of High Energy Physics, Chinese Academy of Sciences,
Beijing 100049, China}

\email{wengss@ihep.ac.cn; zhangsn@ihep.ac.cn}

\begin{abstract}

Low-mass X-ray binaries (LMXBs) are systems in which a low-mass
companion transfers mass via Roche-lobe overflow onto a black hole
(BH) or a weakly magnetized neutron star (NS). It is believed that
both the solid surface and the magnetic field of an NS can affect
the accretion flow and show some observable effects. Using the
disk emission dominant data, we compare the disk evolution of the
two types of systems from low luminosity to super-Eddington
luminosity. As the luminosity decreases the disk in the NS LMXB
4U1608--522 begins to leave the innermost stable circular orbit
(ISCO) at much higher luminosity ($\sim$ 0.1 $L_{\mathrm{Edd}}$),
compared with BH LMXBs at much lower luminosity ($\sim$ 0.03
$L_{\mathrm{Edd}}$), due to the interaction between the NS
magnetosphere and accretion flow. However, as the luminosity
increases above a critical luminosity, the disks in BH and NS
LMXBs trace the same evolutionary pattern, because the
magnetosphere is restricted inside ISCO, and then both the NS
surface emission and (dipole) magnetic field do not significantly
affect the secular evolution of the accretion disk, that is driven
by the increased radiation pressure in the inner region. We
further suggest that the NS surface emission provides additional
information of accretion disk, not available in BH systems.
Through the observed NS surface emission, we argue that the disk
thickness $H/R$ is less than $0.3-0.4$, and that the significant
outflow from inner disk edge exists at luminosity close to
Eddington luminosity.

\end{abstract}


\section{Introduction}

Low-mass X-ray binaries (LMXBs) are systems in which a low-mass
companion ($M < 1M_{\rm \odot}$) transfers mass via Roche-lobe
overflow onto a black-hole (BH) or a weakly magnetized neutron
star (NS; see reviews by Tanaka \& Lewin 1995; van Paradijs \&
McClintock 1995). Based on their X-ray spectral and timing
properties, LMXBs with an NS primary can be classified into two
subclasses (Hasinger \& van der Klis 1989; van der Klis 2006): Z
sources with high luminosity (close to Eddington luminosity,
$L_{\mathrm{Edd}}$) and Atoll sources with low luminosity ($\sim$
0.001--0.5 $L_{\mathrm{Edd}}$).

Compared to NSs, astrophysical BHs are relatively simple objects,
possessing only mass and spin. As luminosity increases, BH LMXBs
go through the quiescent state, low/hard state (LHS), intermediate
state (IMS), high/soft state (HSS), and very high state (VHS)
(Esin et al. 1997). Because IMS and VHS have similar spectral and
timing behavior but with different luminosity, they can be taken
as the same state and represent transitions between LHS and HSS
(Done et al. 2007).

In the X-ray color-color diagram (CCD) or hardness intensity
diagram (HID), Atoll sources trace a U-shape or C-shape track as
the spectra evolve. From top to bottom, the three states are the
extreme island state, the island state, and the banana state
(Gierli\'nski \& Done 2002; Lin et al. 2007). To be consistent
with BH LMXBs, however, we refer to these branches as LHS, IMS and
HSS, respectively. On the other hand, Z sources display an
approximate Z shape in CCD, and the upper, diagonal, and lower
branches are called horizontal, normal, and flaring branches
(HB/NB/FB), respectively. Z sources typically stay at high
luminosity (close to Eddington luminosity) with very soft spectra
on all three branches (Lin et al. 2009, hereafter LRH09).

There are two key differences between NSs and BHs -- the presence
or the absence of a solid surface and a (dipole) magnetic field.
These differences give rise respectively to Type I X-ray bursts,
which are thermonuclear explosions in the surface layers of NSs,
and coherent pulsations, which are the signals resulting from the
dipolar magnetic fields anchored in NSs (Done et al. 2007). These
two phenomena are very common in NS X-ray binaries, but have never
seen in the BH systems. It is believed that both the solid surface
and the magnetic field can also affect the accretion flow, and
show some observable effects.

For comparable mass accretion rates, the observational data
suggest that BH LMXBs are fainter than NS LMXBs by a factor of
$\sim$ 100-1000 in quiescent state. The large X-ray luminosity
difference can be naturally explained by the advection-dominated
accretion flow (ADAF) model. The bulk of thermal energy is trapped
in the advective flow entering into the BH event horizon, and is
lost from sight. Whereas in the case of NS, the thermal energy is
radiated from its solid surface, and makes NS LMXB much brighter
than BH system (Narayan \& McClintock 2008). However, ADAF model
overestimates the luminosity of quiescent NS LMXB, unless most of
the accretion flow is prevented from reaching the NS surface on
account of the ``propeller" effect (Zhang et al. 1998; Menou et
al. 1999).

The interaction of accretion flow with magnetic field can be
characterized by the size of magnetosphere co-rotating with the
central NS; the boundary of the magnetosphere is determined where
the ram pressure of the flow is balanced by the magnetic pressure.
When falling into the magnetosphere, the accreting gas is forced
to co-rotate with the magnetosphere/NS, since the magnetic force
dominates the flow dynamics in this region (Lamb et al. 1973). The
radius of an NS magnetosphere increases with decreasing accretion
rate (ram pressure), that is,
\begin{equation}\label{lamb}
R_{\rm m}=2.7\times 10^8\mbox{ }\left(\frac{L_{\rm
X}}{10^{37}\mbox{ }{\rm erg}\mbox{ }{\rm s}^{-1}}\right)^{-2/7}
\left(\frac{M}{1.4M_{\rm \odot}}\right)^{1/7}
\left(\frac{B}{10^{12}\mbox{ }{\rm G}}\right)^{4/7}
\left(\frac{R_{\rm NS}}{10^6\mbox{ }{\rm cm}}\right)^{10/7}\mbox{
}{\rm cm},
\end{equation}
given by Cui (1997), where $L_{\rm X}$ is the bolometric X-ray
luminosity, $B$ is the NS surface magnetic field strength, and
$R_{\rm NS}$ is the NS radius. If the magnetosphere expands beyond
the co-rotation radius, the centrifugal barrier prevents most
material accreting onto NS, and the disk is truncated at the
magnetosphere radius. It is well known as ``propeller" effect, and
its evidence has been reported in some X-ray pulsars and Atoll
sources (Cui 1997; Zhang et al. 1998; Campana et al 1998). The
Atoll source 4U1608--522 is found to undergo an abrupt spectral
change during the luminosity declines of its 2004 outburst (Chen
et al. 2006). Chen et al. (2006) argued that this event can be
interpreted as the propeller driven spectral state transition,
similar to that found in Aql X-1 (Zhang et al. 1998; Campana et al
1998). In the next section, we investigate the 2007 outburst of
4U1608--522, that was observed with Swift and the Rossi X-ray
Timing Explorer (RXTE) simultaneously. We compare the BH LMXBs
with NS LMXBs from low to super-Eddington luminosity in
\S\ref{res}, and present our conclusions in \S\ref{con}.

\section{The 2007 outburst of 4U1608--522}\label{4U1608}

4U1608--522 is a typical Atoll source (Hasinger \& van der Klis
1989), in which the kHz quasi-periodic oscillations and the Type I
X-ray bursts have been detected (Nakamura et al. 1989; Remillard
\& Morgan 2005). Its distance of 3.6 kpc is estimated from some
bursts that showed photospheric radius expansion (Nakamura et al.
1989). Zhang et al. (1996) first detected its hard X-ray outburst
in low state over 20-100 keV with the Burst and Transient Source
Experiment on the Compton Gamma Ray Observatory. Taking advantage
of the huge database obtained with RXTE, Lin et al. (2007,
hereafter Lin07) evaluated various spectral models for
4U1608--522. However, due to the lack of sensitivity below 2 keV
with RXTE, they could not provide a substantial constraint on the
thermal component, especially in the low luminosity soft and hard
states (Lin07). Thanks to the superior low energy response and
scheduling flexibility of Swift, we can break the new ground
relative to the comprehensive study of 4U1608--522 by Lin07. Here
we present the simultaneous RXTE and Swift observations of
4U1608--522 during its 2007 outburst in this section, and give a
new insight of its accretion disk evolution from high to low
luminosity soft state. Most of the observations from these two
X-ray missions were made within half to one day, and the
observations of 00030791017 (Swift) and 92401-01-11-04 (RXTE) were
made exactly simultaneously on 2007 June 27, as shown in
Figure~\ref{fig1}. In the next two subsections, we describe the
RXTE and Swift data reduction. The spectral modeling results are
shown in \S\ref{model}.

\subsection{RXTE Data Reduction}\label{RXTE}

\begin{figure}
\begin{center}
\includegraphics[scale=0.6]{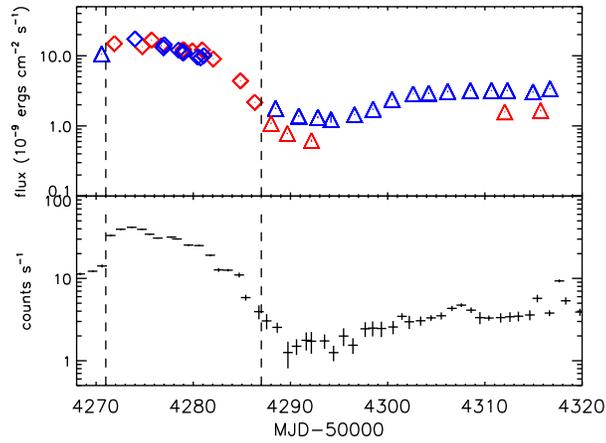}
\caption{\label{fig1} Upper panel: light curves of 4U1608--522
observed with RXTE (blue symbol) at $2.0-50.0$ keV and Swift/XRT
(red symbol) at $0.6-10.0$ keV (see text for description of the
spectral model used). The diamonds between the two dashed lines
correspond to the high/soft state data, and the triangles
correspond to the low/hard state data. The error bars are smaller
than the symbol size. Lower panel: RXTE ASM daily-averaged count
rates at $2.0-12.0$ keV of 4U1608--522 during its 2007 outburst.}
\end{center}
\end{figure}

\begin{figure}
\begin{center}
\includegraphics[scale=0.6]{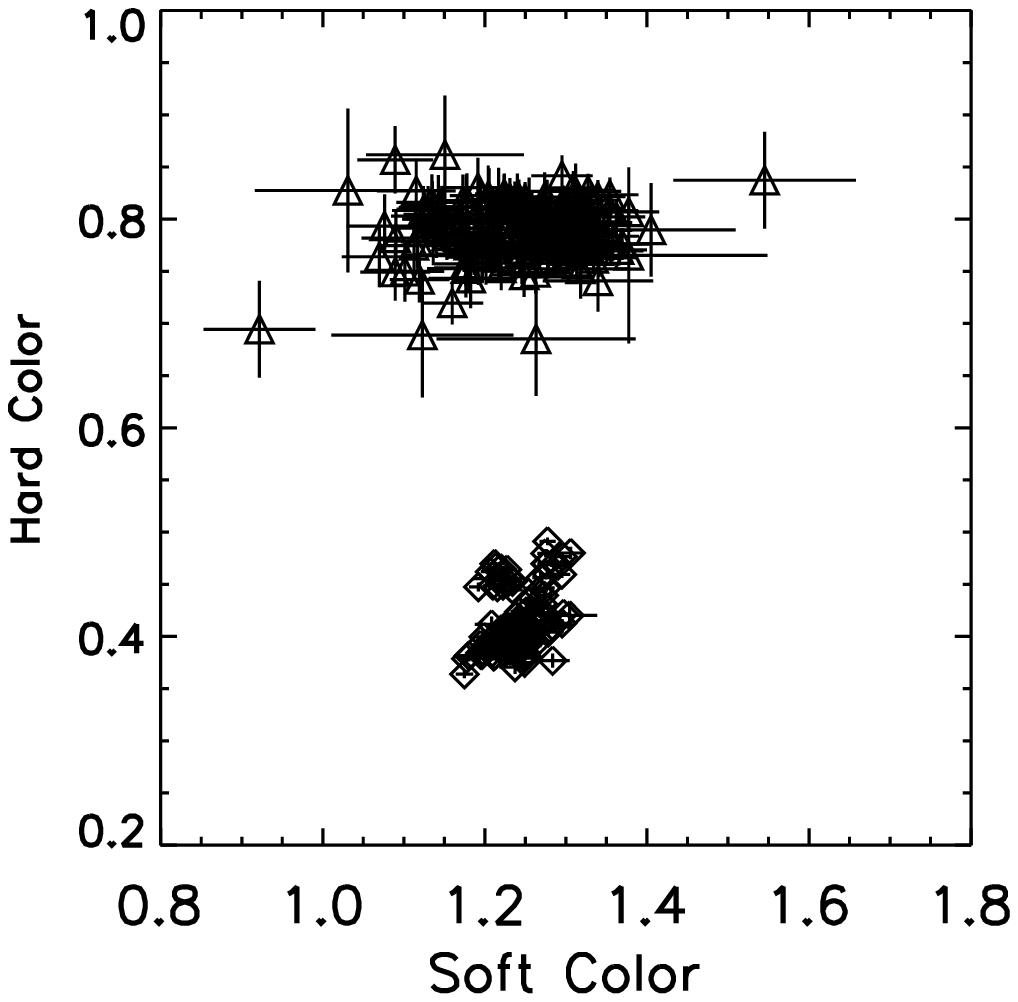}
\caption{\label{fig2} Color-color diagram of 4U1608--522, with bin
size of 128 s. The meaning of the symbols is the same as in
Figure~\ref{fig1}: the diamonds and triangles mark the high/soft
state and low/hard state data, respectively.}
\end{center}
\end{figure}

RXTE has three instruments - the Proportional Counter Array (PCA),
the High-Energy X-ray Timing Experiment, and the All Sky Monitor
(ASM). We analyze 4U1608--522 data from RXTE observations taken
between 2007 June 19 and 2007 August 4 with the FTOOLS software
package version 6.9 \footnote{See
\protect\url{http://heasarc.gsfc.nasa.gov/docs/software/lheasoft/}}.
The detailed studies presented in this work rely on data from the
PCA. We only use the Standard2 data from the top layer of PCU2,
that operated during all the observations (data of 300s before and
1000s after the Type I X-ray burst in July 31 are excluded). The
data are filtered with the standard criteria: the Earth-limb
elevation angle larger than $10\degr$ and the spacecraft pointing
offset less than $0.02\degr$. The background files are created
using the program \texttt{pcabackest} and the latest bright source
background model since the source intensity $>$ 40 counts/s/PCU.
We create the PCA background-subtracted light curves for each
energy channel in 128 s time bins. These light curves are used to
build CCD, defining the soft and hard colors as the ratios of the
counts in the (3.6-4.9)/(2.1-3.6) keV bands and the
(8.6-18.0)/(4.9-8.6) keV bands, respectively. Figures~\ref{fig1}
and ~\ref{fig2} show that the source stays in the lower branch
(diamond, HSS) in the outburst, and traces the upper branch in the
CCD (triangle, LHS) before and after the outburst. However, the
observations are not dense enough to cover IMS.

\subsection{Swift Data Reduction}\label{Swift}

Swift has three scientific instruments on board: the Burst Alert
Telescope (BAT), the X-ray Telescope (XRT), and the UV/Optical
Telescope. 4U1608--522 was observed with XRT from 2007 June 20 to
August 3. The data are reduced and analyzed with the FTOOLS
software package version 6.9.

The XRT data were taken in window-timing (WT) mode before July 12
and after July 30, and switched to photon-counting (PC) mode
between July 12 and July 30. However, the source was so bright
that all PC mode data were strongly affected by pileup. Thus the
PC mode data cannot be used, and only WT mode data are included in
the subsequent analyses.

Initial event cleaning is performed using the \texttt{xrtpipeline}
script, with standard quality cuts, and only events with grades
0-2 are selected as good events. Its spectra are extracted with
\texttt{xselect}, from a rectangular box of 20 pixels wide and 60
pixels long centered at 4U1608--522, and background spectra are
taken from a rectangular region of the same size outside of the
source region. To correct for bad columns, the exposure maps are
generated with \texttt{xrtexpomap}. We also produce the ancillary
response files with \texttt{xrtmkarf} to facilitate subsequent
spectral analyses. The latest response files (v011) are taken from
the CALDB database. Finally, the spectra are grouped to require at
least 20 counts per bin to ensure valid results using $\chi^2$
statistical analysis.

The uncorrected light curves varied drastically because of a
violent change in the XRT effective area (by a factor of $\sim$
10) in the observations on June 22 and June 26. As a result, the
spectra of these two observations cannot be fitted with any
physical model and excluded in the subsequent analyses .

\subsection{Spectral Modeling}\label{model}

Various models are proposed to explain the X-ray spectra of NS
LMXBs (Barret 2001), but not all of them are self-consistent,
though providing acceptable fits. Using multiple evaluation
criteria, Lin07 suggested that the RXTE spectra of NS LMXBs can be
well interpreted with the hybrid model: for the hard state, a
single-temperature blackbody (BB) plus a broken power law (BPL);
for the soft state, two thermal components (standard thin
accretion disk and BB) when the luminosity is high, and an extra
constrained BPL is needed to fit the hard excess above 15 keV due
to low levels of comptonization at lower luminosity. In this work,
we apply their hybrid model to fit the spectra of 4U1608--522 with
XSPEC version 12.6.0 (Arnaud 1996).

The PCA is now well calibrated up to about 50 keV using PCARMF
(v11.7)
\footnote{\protect\url{http://www.universe.nasa.gov/xrays/programs/rxte/pca/doc/rmf/pcarmf-11.7/}},
which was released in 2009 July. The PCA spectra of 4U1608--522
are fitted over the energy range of 2.6-50.0 keV, and a systematic
error of $0.5\%$ is added to all energy channels. On the other
hand, fits made with the XRT data are restricted to the 0.6-10 keV
range to avoid calibration uncertainties at energies less than 0.6
keV (Rykoff et al. 2007).

When the source is in LHS, the PCA spectra are fitted with BB+BPL,
while the Swift/XRT spectra are fitted with BB ({\it bbody} in
XSPEC) plus a power-law (PL, {\it po} in XSPEC) model. 4U1608--522
was active between June 19 and July 4, and the state of the source
transits into HSS. Lin07 showed that the low levels of
comptonization has influence on the spectra above 15 keV and a
constrained BPL ({\it bkn} in XSPEC) component is only required at
low luminosity soft state observations. At higher $L_{\rm X}$,
most spectra can be fit by the multi-color disk model (MCD, {\it
diskbb} in XSPEC) plus BB. Because the RXTE observations do not
cover IMS and the Swift/XRT is lack of hard X-ray data, we fit
both Swift/XRT and PCA spectra with the MCD+BB model. A Gaussian
line with the width fixed at 0.1 keV is added to describe the Fe
line in PCA spectral fitting. Fitting the Swift/XRT spectra, we
find the neutral hydrogen column density of $N_{\rm H}=(0.92-1.06)
\times10^{22}$ ${\rm cm}^{-2}$, which is in good agreement with
those reported in literature (Penninx et al. 1989). Thus all
models include an interstellar absorption component with the
hydrogen column fixed at $N_{\rm H}=1.0\times10^{22}$ ${\rm
cm}^{-2}$ (Lin07), and the unabsorbed flux is calculated in
2.0-50.0 keV for PCA and 0.6-10.0 keV for Swift/XRT, respectively
(Figure~\ref{fig1} and Table 1).

In the soft state, the inner disk radius can be estimated from the
{\it diskbb} model as:
\begin{equation}
R_{\rm in}=\eta \mbox{ } N_{\rm disk}^{0.5} \mbox{ } \frac{D}{10
\mbox{ }{\rm kpc}} \mbox{ }\cos \theta ^{-0.5} \mbox{ } f_{\rm
col}^{2}\mbox{ } {\rm km},
\end{equation}
where $N_{\rm disk}$ is the normalization, $D$ is the distance,
$\theta$ is the angle of the disk, $f_{\rm col}$ is the fractional
change of the color temperature and $\eta$ is the correction
factor for the inner torque-free boundary condition (Zhang et al.
1997; Gierli\'nski \& Done 2002). For X-ray Binaries, we cannot
restrict the inclination very well in most cases, unless the
companion's light curve modulation is observed. Since neither
eclipses nor absorption dips have been observed (Lin07), we assume
a reasonable inclination angle $\sim$ $70\degr$, however, do not
exclude the possibility of smaller ones. $D = 3.6$ kpc, $\eta$ $=$
0.7, and $f_{\rm col}= 1.7$ are adopted in this work. Our main
interest here is the trend of accretion disk evolution that does
not strongly depend on the precise values of the inclination angle
and these correction factors. With the inner disk radius and its
temperature, the bolometric luminosity of the disk can also be
derived as: $L_{\rm disk}=4\pi R^2\sigma_{\rm SB} T^4$. The
Eddington luminosity is $L_{\mathrm{Edd}}=1.3 \times 10^{38}
\times M/M_{\odot}$ erg s$^{-1}$ with $M$ being the mass of
central compact object. Assuming that the mass of NS is 1.4 solar
mass, then its Eddington luminosity is $1.82 \times 10^{38}$ erg
s$^{-1}$ .

\begin{deluxetable}{lllp{0cm}lllp{0cm}}
\tabletypesize{\tiny} \tablewidth{0pt} \tablecaption{Best-fit
Spectral Parameters of Soft State\label{tab:spec}}
\tablehead{\colhead{ObsID} & \colhead{$kT_{\rm disk}$} & \colhead{
$N_{\rm disk}$ } \qquad &\colhead{$kT_{\rm BB}$}\qquad & \colhead{
$N_{\rm BB}$ } \qquad & \colhead{$f_{\rm X}$} &
\colhead{$\chi^2/$dof}} \startdata
&&& RXTE &&&\\
\noalign{\smallskip} \hline \noalign{\smallskip}

92401-01-10-00 &    $1.65_{-0.05}^{+0.05}$  &  $107.8_{-10.1}^{+11.4}$  &  $2.43_{-0.05}^{+0.06}$  &  $0.070_{-0.005}^{+0.005}$  &  $1.741_{-0.006}^{+0.006}$  &  72.39/73 \\
92401-01-10-01 &    $1.54_{-0.06}^{+0.06}$  &  $110.3_{-13.2}^{+15.4}$  &  $2.38_{-0.07}^{+0.08}$  &  $0.051_{-0.005}^{+0.004}$  &  $1.307_{-0.006}^{+0.006}$  &  62.36/73 \\
92401-01-11-00 &    $1.54_{-0.03}^{+0.03}$  &  $111.4_{-6.6}^{+7.0}$    &  $2.41_{-0.02}^{+0.02}$  &  $0.068_{-0.002}^{+0.002}$  &  $1.426_{-0.004}^{+0.004}$  &  68.22/73 \\
92401-01-11-02 &    $1.50_{-0.03}^{+0.03}$  &  $109.8_{-8.5}^{+9.2}$    &  $2.35_{-0.03}^{+0.04}$  &  $0.052_{-0.002}^{+0.002}$  &  $1.198_{-0.004}^{+0.004}$  &  45.56/73 \\
92401-01-11-03 &    $1.42_{-0.09}^{+0.09}$  &  $125.4_{-25.7}^{+32.8}$  &  $2.22_{-0.08}^{+0.10}$  &  $0.052_{-0.006}^{+0.006}$  &  $1.109_{-0.009}^{+0.009}$  &  52.42/73 \\
92401-01-11-04 &    $1.43_{-0.08}^{+0.08}$  &  $128.4_{-23.7}^{+29.8}$  &  $2.31_{-0.09}^{+0.10}$  &  $0.053_{-0.006}^{+0.005}$  &  $1.144_{-0.009}^{+0.010}$  &  51.72/73 \\
92401-01-11-01 &    $1.51_{-0.02}^{+0.02}$  &  $103.2_{-5.7}^{+6.2}$    &  $2.37_{-0.02}^{+0.02}$  &  $0.047_{-0.001}^{+0.001}$  &  $1.115_{-0.003}^{+0.003}$  &  103.7/73 \\
93408-01-01-00 &    $1.43_{-0.03}^{+0.03}$  &  $110.0_{-8.9}^{+9.7}$    &  $2.30_{-0.03}^{+0.04}$  &  $0.041_{-0.001}^{+0.002}$  &  $0.961_{-0.003}^{+0.003}$  &  62.77/73 \\
93408-01-01-01 &    $1.47_{-0.03}^{+0.03}$  &  $97.7_{-8.0}^{+8.8}$     &  $2.35_{-0.04}^{+0.04}$  &  $0.038_{-0.002}^{+0.002}$  &  $0.935_{-0.003}^{+0.003}$  &  88.11/73 \\
93408-01-01-02 &    $1.36_{-0.03}^{+0.03}$  &  $119.8_{-9.2}^{+9.8}$    &  $2.28_{-0.02}^{+0.02}$  &  $0.057_{-0.001}^{+0.001}$  &  $1.001_{-0.003}^{+0.003}$  &  95.24/73 \\

\noalign{\smallskip} \hline \noalign{\smallskip}
&&& Swift/XRT &&&\\
\noalign{\smallskip} \hline \noalign{\smallskip}

00030791010 &   $1.23_{-0.16}^{+0.20}$  &  $217.7_{-76.6}^{+107.6}$  &  $1.83_{-0.21}^{+0.50}$  &  $0.080_{-0.018}^{+0.015}$  &  $1.487_{-0.014}^{+0.014}$  &  692.2/556 \\
00030791013 &   $1.31_{-0.25}^{+0.41}$  &  $143.4_{-76.0}^{+117.0}$  &  $1.92_{-0.27}^{+1.59}$  &  $0.079_{-0.027}^{+0.019}$  &  $1.344_{-0.014}^{+0.015}$  & 669.75/554 \\
00030791014 &   $1.80_{-0.21}^{+0.12}$  &  $69.3_{-12.5}^{+29.2}$    &  $5.37_{-2.68}$          &  $0.210_{-0.116}$           &  $1.674_{-0.020}^{+0.021}$  &  601.503/572 \\
00030791015 &   $1.53_{-0.22}^{+0.09}$  &  $111.1_{-18.7}^{+63.7}$   &  $6.10_{-3.99}$          &  $0.206_{-0.154}$           &  $1.391_{-0.030}^{+0.029}$  & 306.662/312 \\
00030791017 &   $1.34_{-0.14}^{+0.14}$  &  $131.2_{-33.4}^{+47.8}$   &  $2.31_{-0.32}^{+0.63}$  &  $0.077_{-0.006}^{+0.005}$  &  $1.228_{-0.012}^{+0.012}$  & 607.722/587 \\
00030791018 &   $1.28_{-0.14}^{+0.14}$  &  $155.6_{-41.7}^{+62.2}$   &  $2.14_{-0.31}^{+0.65}$  &  $0.064_{-0.006}^{+0.008}$  &  $1.176_{-0.011}^{+0.012}$  & 688.549/567 \\
00030791019 &   $1.48_{-0.12}^{+0.11}$  &  $97.4_{-18.5}^{+26.7}$    &  $3.08_{-0.65}^{+1.66}$  &  $0.086^{+0.064}$           &  $1.214_{-0.012}^{+0.012}$  & 712.975/583 \\
00030791020 &   $1.34_{-0.06}^{+0.06}$  &  $109.1_{-13.4}^{+17.5}$   &  $3.46_{-0.74}^{+1.61}$  &  $0.069_{-0.015}^{+0.059}$  &  $0.894_{-0.008}^{+0.008}$  & 759.609/576 \\
00030791021 &   $0.82_{-0.03}^{+0.03}$  &  $331.2_{-39.4}^{+45.8}$   &  $1.82_{-0.11}^{+0.13}$  &  $0.025_{-0.001}^{+0.001}$  &  $0.438_{-0.004}^{+0.004}$  & 497.757/487 \\
00030791022 &   $0.56_{-0.02}^{+0.02}$  &  $695.9_{-98.1}^{+114.8}$  &  $1.49_{-0.05}^{+0.06}$  &  $0.015_{-0.001}^{+0.001}$  &  $0.217_{-0.003}^{+0.003}$  & 468.277/397 \\

\enddata

\tablecomments{At soft state, the data from PCA are fitted with
{\it wabs*(diskbb+bb+gau)} and the Swift/XRT spectra are fitted
with {\it wabs*(diskbb+bb)}. $kT_{\rm disk}$: inner disk
temperature; $N_{\rm disk}$: normalization of disk; $kT_{\rm BB}$:
BB temperature; $N_{\rm disk}$: normalization of BB; $f_{\rm X}$:
{\bf 2.0--50.0 keV} intrinsic flux in the units $10^{-8}$ erg
cm$^{-2}$ s$^{-1}$ for RXTE, while 0.6-10.0 keV for Swift/XRT;
$\chi^2$/dof: $\chi^2$ and degrees of freedom for the best-fit
model.}

\end{deluxetable}

\begin{figure*}
\begin{center}
\includegraphics[scale=0.7]{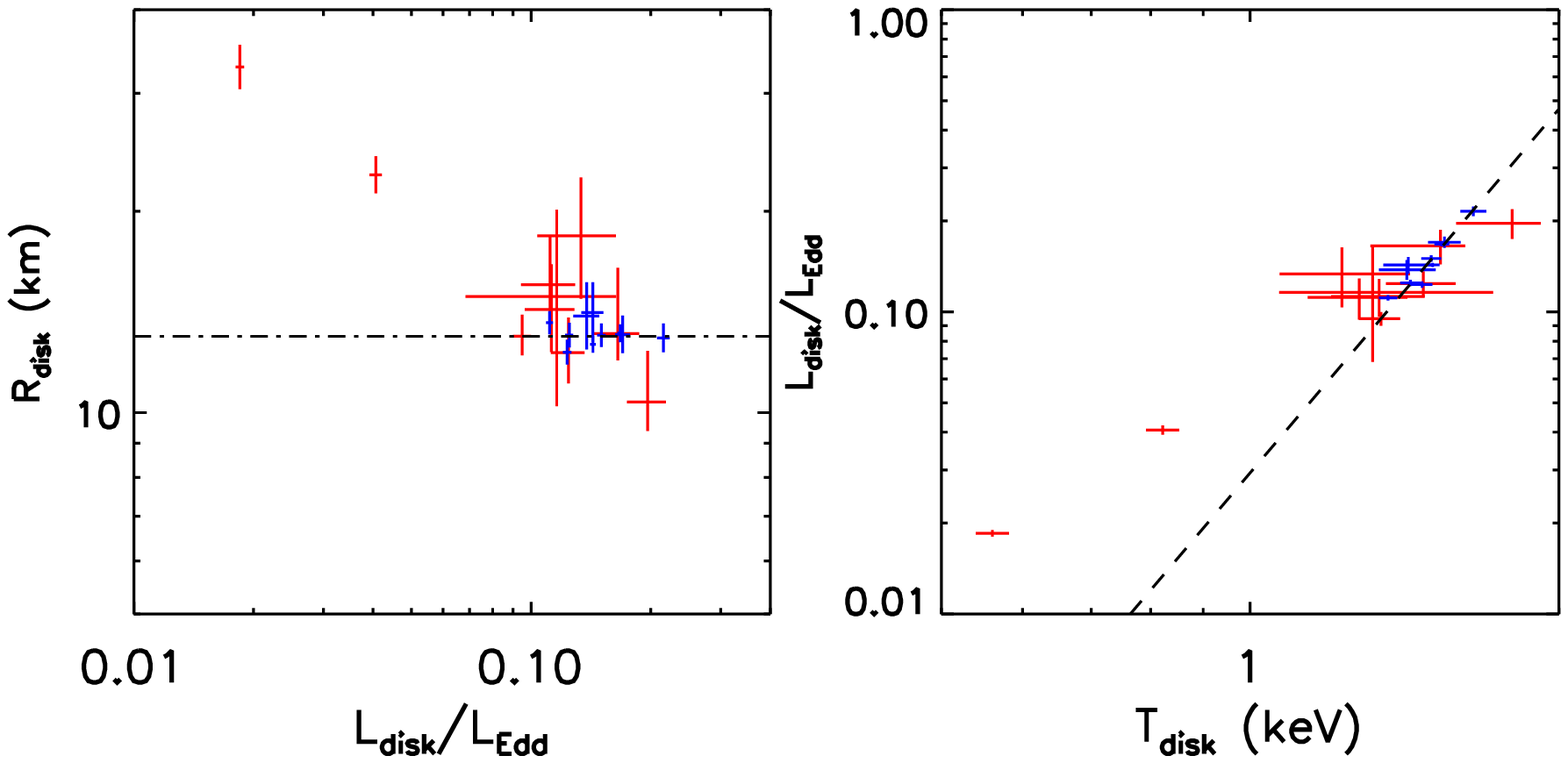}
\caption{\label{fig3} Left panel: the inner disk radius vs. the disk luminosity, and the dot-dashed line
corresponds to $R_{\rm disk} = 13$ km. Right panel: the disk luminosity vs. the inner disk temperature, and the
dashed line shows $L_{\rm disk}=4\pi R^2\sigma_{\rm SB} T^4$ with $R_{\rm disk} = 13$ km. }
\end{center}
\end{figure*}

In Figure~\ref{fig3} we plot the radius of the inner accretion
disk versus the bolometric luminosity of the disk in units of
$L_{\mathrm{Edd}}$. Above $\sim$ 0.1 $L_{\mathrm{Edd}}$, the inner
disk radius remains constant ( $\sim$~13 km), that is, the
innermost stable circular orbit (ISCO) of NS. The dashed line in
the right panel of Figure 3 represents $L_{\rm disk}=4\pi
R^2\sigma_{\rm SB} T^4$ with $R = 13$ km. These results are in
agreement with Lin07 (Figure 7 in their paper).

There are two XRT observations below 0.1 $L_{\mathrm{Edd}}$ that
deviate from the constant radius and $L_{\rm disk} \propto T^4$.
Since these two observations cover the lower luminosity (IMS) just
before the source returns to LHS, we need to check the spectral
model carefully. Fitting the data with MCD+BB and BB+PL shows that
the MCD+BB model is favored; adding a PL component to MCD+BB, the
photon index pegs at the hard limit of 10.0. However, there is a
caveat when we use the simple PL model to depict the Compton
component in fitting the spectra of X-ray binaries: it rises
without limit at low energies, which evidently disagrees with
Comptonization. To eliminate this divergence, we further fit the
data with a more appropriate Compton model (SIMPL, {\it simpl} in
XSPEC), which is developed by Steiner et al. (2009). With only two
free parameters, SIMPL incorporates the basic features of Compton
scattering of soft photons by energetic coronal electrons. Since
the seed photons for the Comptonized component can be from MCD
and/or BB, we combine the SIMPL, DISKBB, and BB in a variety of
ways to fit the spectra (i.e., {\it simpl*(diskbb+bb)}, {\it
(simpl*diskbb+bb)}, and {\it (diskbb+simpl*bb)}). Though all these
models cannot constrain the parameters owing to limited photons,
the fitting results all indicate that the inner disk radius really
moves out.

\begin{figure}
\begin{center}
\includegraphics[scale=0.8]{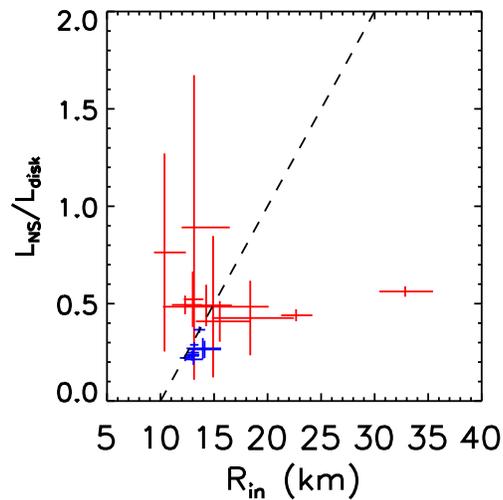}
\caption{\label{fig4} Ratio of BB and MCD luminosity vs. the inner
disk radius.}
\end{center}
\end{figure}

Figure~\ref{fig4} shows the ratio between the BB and MCD
luminosity versus the inner disk radius. Assuming that the
accretion matter falls freely from the inner disk to the surface
of NS, the gravitational potential energy is converted to
radiation on the NS surface. The dashed line represents that the
luminosity of BB component (i.e., the emission from the NS
surface) is equal to the gravitational potential energy, taking NS
radius $R_{\rm NS} = 10$ km. The data above the line mean that the
accretion matter does not fall freely from the disk but also
contains some initial kinetic energy as the matter leaves the
inner disk boundary. The two observational data in IMS lying far
below the line indicate that significant outflow is accompanied
with accretion.

\section{Comparing NS LMXBs with BH LMXBs}\label{res}

\subsection{Transition from Low/Hard State to High/Soft State}\label{ls}

Since Cui (1997) first presented evidence for ``propeller" effect
in two X-ray pulsars, GX 1+4 and GRO J1744-28, further
observational evidence for the interaction between magnetosphere
and accretion disk has been found in other NS X-ray binaries
(e.g., Zhang et al. 1998; Campana et al. 1998). Although the
details vary, all magnetosphere models predict that the inner disk
is truncated at the magnetosphere radius, that expands with
decreasing accretion rate. The entire process should simply
reverse: with increasing accretion rate the magnetosphere moves
inward (Cui 1997). However, when the inner disk is truncated at
several times of $R_{\rm ISCO}$, most of disk components with the
temperature around 0.5 keV are out of the observed bandpass of
RXTE. Thus, direct measurement of magnetosphere radius cannot be
made unless the data from instruments covering the softer X-ray
bandpass are used (Done et al. 2007; Gierli\'nski et al. 2008).

Investigating the simultaneous RXTE and Swift observations of
4U1608--522 during its 2007 outburst, we find that the inner disk
radius increases along the outburst declining phase. This can be
naturally interpreted as the interaction between the magnetosphere
and the accretion disk. Using Equation (1) we can further derive
the magnetic field of $\sim$ $10^{8}$ G, in agreement with the
value derived from the ``propeller" effect (Chen et al. 2006). The
fitting result also implies that magnetically driven outflowing is
significant at lower luminosity ($<$ 0.1 $L_{\mathrm{Edd}}$; see
discussion above about Figure 4), possibly as a result of
disk-magnetosphere interaction.

\begin{figure}
\begin{center}
\includegraphics[scale=0.8]{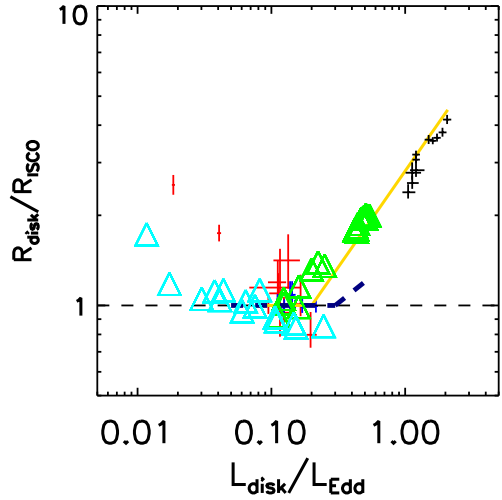}
\caption{\label{fig5} The inner disk radius normalized to ISCO, as
a function of the disk luminosity, where all data used here are
disk dominant. The red and blue crosses mark the Swift and RXTE
data of 4U1608--522, respectively. We also mark the data of XTE
J1817-330 (Gierli\'nski et al. 2008), GRS 1915+105 (McClintock et
al. 2006), XTE~J1701--462 (LRH09), Cir X-1 (Ding et al. 2011, in
preparation), and LMC X-3 (Steiner et al. 2010) by cyan triangles,
green triangles, the gold line, black crosses, and the navy dashed
line respectively. Please see the text for the mass of central
compact object $M$, and the Eddington luminosity
$L_{\mathrm{Edd}}=1.3 \times 10^{38} \times M/M_{\odot} $ erg
s$^{-1}$ is used here.}
\end{center}
\end{figure}

To explain the spectral state of BH LMXBs, Esin et al. (1997)
proposed a possible scenario: an accretion flow around a BH
consists of two zones with a transition radius $R_{\rm tr}$, an
inner ADAF, and an outer standard thin disk. The ADAF can extend
outside, lying above the disk in the form of a hot corona. The
transition radius is correlated inversely with accretion rate,
that is, the outer thin disk is restricted to a larger radius at a
lower mass accretion rate, and vice versa. Substantial theoretical
and observational efforts have been made since then; however, some
major issues still remain unresolved so far. More observations
have shown that the transition occurs at a range of luminosity,
even in a single object. These suggest that the mass accretion
rate is a dominant but not the only parameter in determining
spectral state transitions (Yu et al. 2004). Because the behavior
of a disk is hard to track as the temperature is below 0.5 keV and
its contribution is dominated by non-thermal component, whether
the thin disk extends down close to the central compact object in
the hard state has remained another controversial topic for a long
time (Liu et al. 2011). Using the same data of XTE~J1817--330
during its 2006 outburst, Rykoff et al. (2007) showed that the
cool disk remained near ISCO at very low luminosity $\sim$ 0.001
$L_{\mathrm{Edd}}$, but Gierli\'nski et al. (2008) argued that the
disk receded when the source left the disk dominant soft state. To
avoid this uncertainty, we only investigate the disk dominant
observations in the present work. It was found that its inner disk
radius increases as the luminosity falls below $\sim$ 0.02-0.03
$L_{\mathrm{Edd}}$ (cyan triangles in Figure~\ref{fig5}) (Rykoff
et al. 2007; Gierli\'nski et al. 2008); the mass of BH in
XTE~J1817--330 is 6 solar masses and the inclination angle is
$60\degr$ (Sala et al. 2007). Some other work also supported that
the disk in BH LMXBs begins to leave ISCO as the luminosity
decreases below $\sim$ 0.02-0.03 $L_{\mathrm{Edd}}$ or even lower
(Nowak et al. 2008; Maccarone 2003).

When the accretion rate is sufficiently high in NS LMXB system,
the magnetosphere can be pushed inside ISCO, and thus
significantly decoupled from the accretion disk, assuming that the
disk's physical boundary cannot extend beyond ISCO. As a result,
above this critical luminosity, the disk in BH and NS LMXBs should
share very similar evolutionary pattern, as the disk-magnetosphere
interaction in NS LMXBs is not important in this case. Figure 5
shows that the thin accretion disk extends down to ISCO for both
BH LMXBs and Atoll sources in HSS with luminosity $\sim$ 0.1-0.3
$L_{\mathrm{Edd}}$. It has been suggested that ISCO is just
outside the radius of NS, and providing constraint on the NS
equation of state (Done et al. 2007). On the other hand, ISCO is a
monotonically decreasing function only of the BH spin. Thus, the
spin of BH can be immediately derived from the ISCO if the
distance, mass, and inclination of the systems are available
(Zhang et al. 1997; for recent developments please refer to
McClintock et al. 2006).

\subsection{Transition from High/Soft State to Super-Eddington State}\label{super}

It has been realized for a long time that the super-Eddington
accretion flow seems to occur in many astrophysical situations,
such as in ultra-luminous X-ray sources (ULXs), Narrow Line
Seyfert 1's, and the growth of the first BHs in the early universe
(Done et al. 2007). Investigating the spectra of M33 X-8, Weng et
al. (2009) suggested that some of ULXs occupy a new state --
super-Eddington state (or slim state; see also Gladstone et al.
2009). However, because all these sources are out of the Galaxy
and thus with inadequate data, and by contrast, most BH LMXBs and
Atoll sources in the Galaxy are well below the Eddington
luminosity, our knowledge on the super-Eddington accretion flow is
extremely rudimentary currently.

GRS 1915+105 is one of the few BH LMXBs that transit among the
five classical states and super-Eddington state. Fitting the
thermal dominant data of GRS 1915+105, McClintock et al. (2006)
found that the value of the BH's spin remained constant when its
luminosity was less than 0.3 $L_{\mathrm{Edd}}$, but was depressed
when the luminosity exceeded a critical level. Observationally it
means that the disk inner edge increased with luminosity. Such
phenomenon can be explained in two very different ways. (1) The
inner disk edge is still located at ISCO, but its emission is
shaded by the outer disk when the luminosity goes higher and the
disk becomes thicker. Therefore the inner disk radius obtained
from the spectral fitting is not the true value but just the
larger radius at which the emission is not blocked by outer disk
(McClintock et al. 2006). (2) The inner disk really recedes at
high luminosity. However, the data of GRS 1915+105 alone cannot
distinguish between these two pictures.

Unlike Atoll sources, Z sources typically radiate at luminosity
close to Eddington luminosity. During its 2006-2007 outburst, the
extraordinary NS LMXB XTE~J1701--462 evolved from super-Eddington
luminosity to quiescence. The source firstly displays Cyg-like Z
track, then Sco-like Z track, finally evolved smoothly into the
Atoll track (see LRH09 for the definition of the source branches),
with the disappearance of the FB and through the NB/FB vertex.
According to the spectral analysis, the intimate relation between
the Atoll track and the NB/FB vertex is suggested by LRH09. The
Atoll HSS is characterized by a constant inner disk radius,
whereas the NB/FB vertex stage exhibits a luminosity dependent
expansion of the inner disk (LRH09).

We plot the ISCO-scaled radius of inner disk versus the luminosity
in units of $L_{\mathrm{Edd}}$ in Figure 5, where $D = 11.0$ kpc,
$M = 14.0$ $M_{\rm \odot}$, and $\theta = 66\degr$ are adopted for
GRS 1915+105, and the inner disk radii obtained from the Figure 1.
in McClintock et al. (2006) are marked by green triangles. The
inner disk radii of XTE~J1701--462 taken from Figure 17 in LRH09
are denoted by the gold solid line, because the number of data
points is too numerous for display. Note in LRH09,
$L_{\mathrm{Edd}}=3.79 \times 10^{38}$ erg s$^{-1}$ is adopted for
NS, whereas $L_{\mathrm{Edd}}=1.82 \times 10^{38}$ ergs s$^{-1}$
is used here.

It is interesting that the disk evolution of GRS 1915+105 and
XTE~J1701--462 track the same way from HSS to super-Eddington
state as shown in Figure~\ref{fig5}. This implies that both NS
surface emission and (dipole) magnetic field do not significantly
affect the secular evolution of the accretion disk. Note that such
evolutionary feature is also found in other Z sources, e.g. Cir
X-1 (Ding et al. 2011, in preparation) and BH X-ray binaries, e.g.
LMC X-3 (Steiner et al. 2010). Steiner et al. (2010) analyzed
hundreds of observations of LMC X-3 collected by eight X-ray
missions, and found that the source is habitually soft and highly
variable. The inner disk radius stably kept at ISCO when LMC X-3
varied between low Eddington luminosity ($\lesssim$ 0.05
$L_{\mathrm{Edd}}$) and $\sim$ 0.3 $L_{\mathrm{Edd}}$, whereas it
moved out when the source is brighter than 0.3 $L_{\mathrm{Edd}}$
(see Figure 2 in Steiner et al. 2010). Thus this
luminosity-dependent expansion of the inner disk cannot be
attributed to the extreme physical parameters in GRS 1915+105 or
XTE~J1701--462 (e.g., high spin in GRS 1915+105), but would be
caused by the radiation pressure enhancement at luminosity close
to Eddington luminosity (i.e., the local Eddington limit effect as
mentioned in LRH09).

Besides the evolution of the accretion disk, Figure 17 in LRH09
showed that the boundary emission area maintained its small and
nearly constant size from the Atoll stage to the NB/FB vertex.
Lin07 speculated that the small size of the boundary layer can be
evidence of a geometrically thin accretion stream feeding a rather
well-defined impact zone, and the BB emission is restricted in a
small symmetric equatorial belt on the NS surface. We would like
to further suggest that, the quantitative constraint of disk
thickness can be given by the result of the constant BB radius if
the inclination of system is known. Assuming that a disk thickens
with higher luminosity, when the value of thickness $H/R$ $>$
$\cos \theta \sim 0.3-0.4$, the boundary layer emission should be
partially obscured by the outer disk, and the observed BB radius
should decrease with intensity increase. However, this picture
obviously conflicts the observed constant BB radius. In the NB/FB
vertex branch, the luminosity of BB is much less than the disk
luminosity, and the deficit of the BB component becomes more
severe when luminosity increases, implying that most inflowing
matter is ejected as outflow before reaching the NS surface.

\section{Conclusions}\label{con}

Type I X-ray bursts and the coherent pulsations are unique
signatures of NSs; however, not all NSs show these characteristics
(Done et al. 2007; Lewin et al. 1993). It is believed that both
the solid surface and the magnetic field can affect the accretion
flow and show some observable effects as evidence of NSs.
Comparing the disk evolution of BH and NS X-ray binary systems
with the disk dominant data, we find that different phenomena
exist below $\sim$0.1 $L_{\mathrm{Edd}}$, whereas the same
evolution pattern is shared above the critical luminosity. Our
main results and discussions are summarized in the following.

1. The inner disk expands outward at low luminosity as luminosity
decreases in the soft state in both BH and NS LMXBs' systems. The
disk in a BH LMXB stays at ISCO until luminosity near or even
below $\sim$ 0.03 $L_{\mathrm{Edd}}$ due to the disk's secular
evolution. However, the disk in the Atoll source 4U1608--522
begins to leave ISCO at much higher luminosity $\sim$ 0.1
$L_{\mathrm{Edd}}$, that can be interpreted as the interaction
between magnetosphere and accretion flow, providing further
evidence for NS in the system. The fitting result also indicates
that the magnetically driven outflow is significant at lower
luminosity and the NS's surface (dipole) magnetic field is $\sim$
$10^{8}$ G in 4U1608--522, in good agreement with previous works.
However, a major drawback of the work presented here is the small
sample of sources investigated, and thus at this stage we cannot
claim with confidence generality of the results presented here,
that should be confronted in the future with further analyses of
more observations in larger samples.

2. Above the critical luminosity, the disks in BH and NS LMXBs
trace the same evolutionary pattern. We suggest that the
magnetosphere is pushed inside ISCO, and both NS surface emission
and (dipole) magnetic field do not significantly affect the
secular evolution of the accretion disk at high luminosity. When
the source is brighter than 0.3 $L_{\mathrm{Edd}}$, its disk
expands roughly as $R_{\rm disk} \propto L_{\rm disk}$, as a
result of radiation pressure increase in the inner region. The
simple standard thin disk model rather than more sophisticated
models is used in this work, because of its simplicity, uniform
treatment of all spectral fittings, and consistent trend of disk
evolution obtained. However, some physical assumptions of the thin
disk would break down at high luminosity. To more accurately
describe the high accretion rate flow, relativistic effects,
self-irradiation of the disk, color correction factor, the inner
torques on the disk, etc., should be taken into account and
treated with caution in the future.

3. We have made the first quantitative constraint on the disk
thickness, $H/R$ $<$ $0.3-0.4$, with the observed surface NS BB
emission at around Eddington luminosity; our result also implies
that most inflowing matter is ejected as outflow before reaching
the NS surface. Since we expect at high luminosity the accretion
disks in BH systems have the same property as that in NS systems,
the above conclusion indicates that the advected fraction should
be very small in BH systems at high accretion rate, because most
accreted material should be ejected as outflow in this case.

Finally we mention that we only studied in this work the Atoll
HSS, the NB/FB vertex branch of XTE J1701--462, and the disk
dominant state of GRS 1915+105. However, both two sources are
variable, and the spectra show complicated features along their
transitions. Different branches may contain very different
accretion physics (LRH09; Fender \& Belloni 2004), that is beyond
the scope of this work.

 \acknowledgments{}
This research has made use of data obtained from the High Energy
Astrophysics Science Archive Research Center (HEASARC), provided
by NASA's Goddard Space Flight Center.

We thank the anonymous referee for his/her constructive criticism
and suggestions, that have allowed us to improve significantly the
presentation of this paper.  S.S.W. thanks Dr. Weimin Gu and Dr.
Guoqiang Ding for helps on data analysis and many interesting
discussions. S.N.Z. acknowledges partial funding support from the
Directional Research Project of the Chinese Academy of Sciences
under project no. KJCX2-YW-T03 and by the National Natural Science
Foundation of China under grant nos. 10821061, 10733010, 10725313,
and by 973 Program of China under grant 2009CB824800.

\end{document}